\RequirePackage{rotating}
\documentclass[usenatbib]{mnras}
\usepackage{epsfig}
\usepackage{amsmath}
\usepackage{graphicx}
\usepackage{array}
\usepackage{textcomp}
\usepackage{amssymb}
\usepackage{rotating}
\usepackage{pdflscape}
\usepackage{caption}
\usepackage{subcaption}
\usepackage{longtable}
\usepackage{xtab}
\def\mnras {{\sc MNRAS}}
\def\apj {ApJ}
\def\apjs {ApJS}
\def\apjl {ApJL}
\def\aj {AJ}

\def\pasp {PASP}
\def\pasj {PASJ}
\def\aap {A\&A} 
\def\nat {Nature}

\title[Bars in MaNGA]
      {SDSS-IV MaNGA: Spatially resolved star formation in barred galaxies} 
\author[A.\ Fraser-McKelvie et al.]
       {Amelia Fraser-McKelvie$^{1}$\thanks{Amelia.Fraser-McKelvie@nottingham.ac.uk}, Alfonso Arag\'on-Salamanca$^{1}$, Michael Merrifield$^{1}$, \and Karen Masters$^{2}$, Preethi Nair$^{3}$, Eric Emsellem$^{4}$, Katarina Kraljic$^{5}$, \and Dhanesh Krishnarao$^{6}$, Brett H. Andrews$^{7}$, Niv Drory$^{8}$, Justus Neumann$^{9}$
                \vspace*{1mm}\\
        $^{1}$ School of Physics \& Astronomy, University of Nottingham, University Park, Nottingham, NG7 2RD, U.K. \\
        $^{2}$ Department of Physics and Astronomy, Haverford College, 370 Lancaster Ave, Haverford, PA 19041, USA \\
        $^{3}$ Department of Physics and Astronomy, University of Alabama, Tuscaloosa, AL 35487, USA \\
       $^{4}$ European Southern Observatory, Karl-Schwarzschild-Str. 2, 85748 Garching, Germany \\
        $^{5}$ Institute for Astronomy, University of Edinburgh, Royal Observatory, Blackford Hill, Edinburgh EH9 3HJ, U.K. \\
        $^{6}$ Department of Astronomy, University of Wisconsin-Madison, 475N. Charter St., Madison WI 53703, USA\\
        $^{7}$ PITT PACC, Department of Physics and Astronomy, University of Pittsburgh, Pittsburgh, PA 15260, USA\\
        $^{8}$ McDonald Observatory, The University of Texas at Austin, 1 University Station, Austin, TX 78712, USA\\
         $^{9}$ Institute of Cosmology \& Gravitation, University of Portsmouth, Dennis Sciama Building, Portsmouth, PO1 3FX, U.K. \\
	}
\begin{document}
\maketitle
\begin{abstract}
Bars inhabit the majority of local-Universe disk galaxies and may be important drivers of galaxy evolution through the redistribution of gas and angular momentum within disks. 
We investigate the star formation and gas properties of bars in galaxies spanning a wide range of masses, environments, and star formation rates using the MaNGA galaxy survey. 
Using a robustly-defined sample of 684 barred galaxies, we find that fractional (or scaled) bar length correlates with the host's offset from the star-formation main sequence. 
Considering the morphology of the H$\alpha$ emission we separate barred galaxies into different categories, including barred, ringed, and central configurations, together with H$\alpha$ detected at the ends of a bar. 
We find that only low-mass galaxies host star formation along their bars, and that this is located predominantly at the leading edge of the bar itself. 
Our results are supported by recent simulations of massive galaxies, which show that the position of star formation within a bar is regulated by a combination of shear forces, turbulence and gas flows. We conclude that the physical properties of a bar are mostly governed by the existing stellar mass of the host galaxy, but that they also play an important role in the galaxy's ongoing star formation.

\end{abstract}

\section{Introduction}
Bars inhabit most disk galaxies in the present-day Universe \citep[e.g.][]{deVauc63a, Eskridge00, Knapen00, Nair10, Masters11}. Their prevalence means they are clearly important structures in these systems, but do they play a wider role in galaxy evolution? The multiple (and sometimes contrary) observed effects of a bar on its host galaxy suggest complex physical processes are at play that are yet to be described in a satisfactory manner.

Bars are postulated to both enhance and suppress star formation in their host galaxies. That said, it is difficult to disentangle whether a bar is the driver behind, or the result of, the cessation of star formation in a galaxy. Possible observational evidence for the involvement of a bar in star formation cessation includes the fact that barred galaxies are consistently redder than their unbarred counterparts, and gas fractions and star formation rates are lower at fixed mass \citep[e.g.][]{Masters12, Wang12, Kruk18}, though some studies find no evidence for this \citep{Erwin18, Diaz-Garcia19}. A high fraction of passive spiral galaxies also host bars \citep{Fraser-McKelvie18}. 

Physically, large-scale bulk motions of gas are preferentially funnelled along a bar, some of which is deposited onto central regions, whilst simultaneously starving the inner disk of the fuel for star formation \citep[e.g.][]{Sellwood93}. 
The gas deposited in the central regions of a galaxy via the bar may be used up in a burst of star formation \citep[e.g.][]{Ho97, Coelho11}, feed the central black hole \citep{Shlosman89,Jogee06}, or build up the central mass concentration \citep{Kormendy04}. It is not clear whether the presence of a bar is the main driver for  rapid gas consumption, but this evidence suggests that bars are at least strongly linked with this process.

Evidence for enhanced star formation in the centres of barred galaxies also exists \citep[e.g.][]{Heckman80, Knapen95}, often in the form of nuclear rings \citep[e.g.][]{Comeron10}. The timescale for this star formation is unknown, though predicted to be short (and possibly a series of sustained bursts) from stellar population analyses of the nuclear ring regions \citep[e.g.][]{Allard06}. A period of short-lived star formation in a bar is also supported by \citet{Ellison11}, who find a central metal enhancement but no corresponding star formation rate (SFR) enhancement in fibre measurements of low-mass, barred galaxies at low redshift. It may be that the role of bars in the enhancement of star formation was more significant in the past \citep[e.g.][]{Carles16}. 

The growth of bars within disks is not understood in detail. 
Bars generally grow via the capture of existing disk stars, though they can also produce new stars from funnelled gas.
The balance between bar growth and disk growth must be critical: simulations show too much gas funnelling can also destroy bars \citep{Bournaud02}. The majority of bars will form from a disk instability, and the more gas-poor the disk is the easier it is to form a bar, leading us to imply that bars in more passive galaxies were in place earlier \citep{Sheth08,Athanassoula13}. If we accept that bars grow bigger with time, then bar length, or indeed strength \citep{Kim17} could be a good indicator of bar evolutionary stage.

Whatever the result on the global SFR, there is a good deal of evidence for gas flow along bars, including higher central molecular gas content in barred vs. unbarred spirals \citep{Sakamoto99,Sheth05}, higher central metallicities in barred galaxies \citep{Ellison11}, and central holes in the HI maps of strongly-barred galaxies \citep{Newnham19}.
Gas flows are thought to initially accelerate circumnuclear star formation, before contributing to the overall quenching of a galaxy. Observations of an anti-correlation between the likelihood of a galaxy hosting a strong bar and its specific star formation rate support this theory \citep[e.g.][]{Gavazzi15}. 
In addition, a bi-modality between bar likelihood and bar length with bulge prominence suggests that the growth of disky pseudobulges may be a side-effect of bar evolution \citep{Cheung13}.
Catching a bar in the act of funnelling gas is rare -- either because this phenomenon is short lived \citep{Bournaud05}, happens only in a small fraction of galaxies \citep{Verley07}, or the net inflow rate is so small that it is difficult to observe until recently without targeted studies of individual galaxies and excellent spatial resolution \citep[e.g.][]{Hunt08, Holmes15}. 
Ionised gas flow along a bar is even rarer, though previous studies of small samples of galaxies have observed the streaming of [NII] and H$\alpha$ \citep[e.g.][]{deVauc63, Zurita04,Pan15}. This ionised gas flow is the smoking gun of star formation occurring within bars.

Given the plethora of gas and dynamical processes occurring within barred galaxies, previous studies with small sample sizes have been too limited to begin to disentangle all of the potential processes occurring from each other.
What has been missing from the literature is a large sample of barred galaxies for which spatially-resolved spectroscopy is available. In this paper, we employ the Mapping Nearby Galaxies at APO \citep[MaNGA;][]{Bundy15} galaxy survey, which provides a larger, more well-defined sample of barred galaxies with a wide ranges of stellar masses, environments, bar morphologies, star formation rates, and H$\alpha$ morphologies extracted from the data cubes.
The wealth of data from MaNGA and its ancillary programs will allow us to study these systems statistically, to determine how bars fit into the wider evolutionary picture of their host galaxies.
In this paper we use a flat $\Lambda$CDM cosmology with $H_{0}=70~\rm{km}~\rm{s}^{-1}~\rm{Mpc}^{-1}$ $h=H_{0}/100$, $\Omega_{M}=0.3$, $\Omega_{\Lambda}=0.7$, and a \citet{Kroupa02} IMF.

\section{Data}
\subsection{The MaNGA Galaxy Survey}
The MaNGA Galaxy Survey is an integral field spectroscopic survey that will observe 10,000 galaxies \citep{Bundy15, Drory15} by survey completion. It is an SDSS-IV project \citep{Blanton17}, employing the 2.5m telescope at Apache Point Observatory \citep{Gunn06} and BOSS spectrographs \citep{Smee13}. MaNGA's target galaxies were chosen to include a wide range of galaxy masses and colours, over the redshift range $0.01<z<0.15$. The Primary+ sample \citep[][]{Yan16b, Wake17} contains galaxies with spatial coverage out to $\sim$1.5 $R_{\textrm{e}}$ for $\sim$66\% of the total sample, and the remainder (dubbed the Secondary sample) are observed out to  $\sim$2.5 $R_{\textrm{e}}$, generally at higher redshifts than the Primary+ sample. 

MaNGA Product Launch 8 (MPL-8) contains 6779 unique galaxy observations, observed and reduced by the MaNGA Data Reduction Pipeline \citep{Law15}, with derived properties produced by the MaNGA Data Analysis Pipeline \citep[DAP;][]{Westfall19}, provided as a single data cube per galaxy \citep{Yan16}. 

\subsection{Barred Galaxy Sample Selection}
We select a sample of barred galaxies from the MaNGA survey using Galaxy Zoo 2 \citep{Willett13}. Galaxy Zoo 2 was a citizen science project that asked participants to classify galaxies according to a flow chart of questions about a galaxy's morphology. Based on user identifications (and weighting individual scorers on their accuracy), a probability that a galaxy contained a particular feature was derived. To account for user error, we employ weighted fraction values and find through trial and error that the optimal combination to select barred galaxies is:
\begin{itemize}
\item $\texttt{p\_bar\_weighted > 0.5}$
\item $\texttt{p\_not\_edgeon > 0.5}$
\end{itemize}

This combination of parameters is most effective at both selecting barred galaxies and filtering out edge-on galaxies which users frequently classify as bars. Similar barred sample selections have been made in \citet{Masters12} and \citet{Kruk18}, though we note that the $\texttt{p\_bar\_weighted > 0.5}$ cut used in this work is quite stringent. We decided on this value to prioritise a clean sample over a complete sample and minimise contamination from non-barred galaxies. Given this, and the fact that these are optical images, bar classification will be biased more towards stronger bars \citep{Masters12}, so, the weakest bars may be missing from this selection. From the starting sample of 6779 galaxies, we find 684 barred galaxies through this method. While this may seem like a small fraction of the MaNGA sample, we note that a galaxy must be relatively face-on for a user to be able to classify whether a bar is present (indeed, the mean axis ratio of the sample is 0.72). Our sample spans a wide stellar mass ($\rm{M}_{\star}$) range, from $2.0\times10^{8}~\rm{M}_{\odot} - 1.5\times10^{11}~\rm{M}_{\odot}$.

\subsection{Stellar Masses and Star Formation Rates}
Stellar masses are adopted from the NASA-Sloan Atlas \citep[NSA;][]{Blanton11}. The NSA is a reanalysis of SDSS photometry that incorporates better sky subtraction and deblending, which particularly aids in the analysis of larger galaxies. The elliptical Petrosian photometry, along with an increase in redshift range, was added originally for the targeting catalogue of MaNGA. SDSS Data Release 13 contains the new version of the NSA, \texttt{v1\_0\_1}, which consists of 641,409 bright, nearby galaxies. This catalogue also contains measurements of the $r$-band elliptical Petrosian half-light radius, which we will refer to as the effective radius of the galaxy, $R_{\rm{e}}$.

We also determine the integrated star formation rate (SFR) from mid-infrared data provided by the \textit{Wide-Field Survey Explorer} ($WISE$) satellite AllWISE source catalogue \citep{Cutri14}. The $WISE$ 12$\mu$m (W3) and 22$\mu$m (W4) bands are excellent tracers of the interstellar medium (ISM) emission produced by dust heated by star formation. Any residual contamination from old, evolved stellar populations is removed using the SFR calibration of \citet{Cluver17}, which models and subtracts stellar contribution in the W3 and W4 bands before converting this flux to a SFR.

\begin{figure*}
\label{bar_len_sect}
\centering
\begin{subfigure}{0.99\textwidth}
\includegraphics[trim={4cm 0 4cm 0},clip, width=\textwidth]{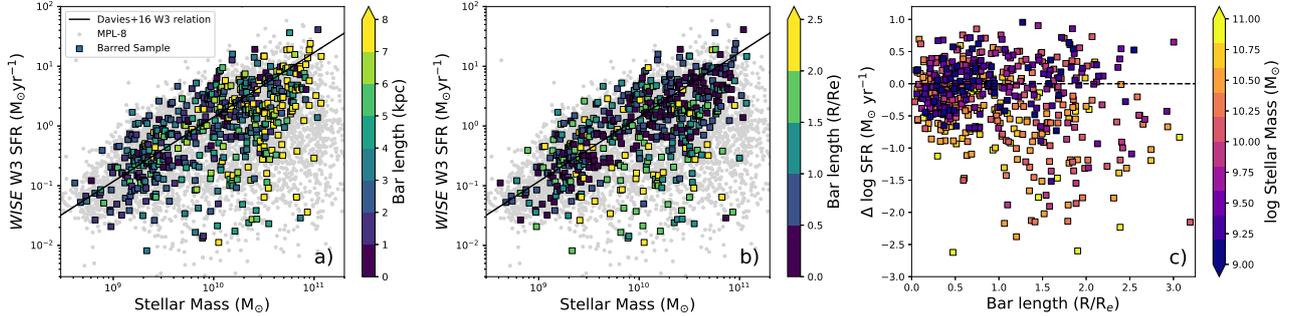}  
\end{subfigure}
\caption{Measures of bar length as a function of galaxy stellar mass, star formation rate, and distance from the SFMS line. Panel a) \& b): star formation rate vs. stellar mass plots of barred MaNGA galaxies, with main sequence line from \citet{Davies16} shown in black. Grey points are all galaxies in MaNGA MPL-8, and squares overlaid are the barred sample, colour-coded by bar length in units of kpc (panel a) and $R/R_{\rm{e}}$ (panel b). Panel c): The log vertical distance from the SFMS line for galaxies as a function of  scaled bar length and colour-coded by stellar mass. A negative value of $\Delta \log \rm{SFR}$ indicates a galaxy lies below the SFMS line. Galaxies lying below the main sequence line are more massive than those close to the line for a given bar length.}
\label{bar_length}
\end{figure*}

\section{Measuring Bar Lengths}

One of the most fundamental measurable properties of a bar is its length, and we calculate this for the entire galaxy sample.
We measure bar length using the Fast Fourier Transform bar analysis method of \citet{Kraljic12}, adapted for use with MaNGA datacubes. 
In this approach, the presence of a bar is inferred by a constant value for the phase of the m=2 Fourier mode. The length is calculated by measuring the phase of the second mode, $\Phi_{2}(r)$, within the bar region. We define a bar to be present when $\Phi_{2}(r)$ is constant to within 5\textdegree, and the radii for which this condition is met correspond to radii at which a bar is present. To improve the starting guess, we include a bulge size estimate, derived from bulge-to-total ratios from the \citep{Simard11} catalogue. The length, strength, and angle of the bar are measured for each MaNGA collapsed data cube, which is treated as a white-light image. We note that higher-resolution optical images could have been used for this same analysis, though given we perform the same technique on individual wavelength slices in Section~\ref{bar_angles}, for comparability, we chose to employ the collapsed data cube images. We also note the discrepancies present between bar length measurements techniques. \citet{Diaz-Garcia16a} for example, point out that Fourier techniques such as that employed here typically result in smaller bar lengths than direct visual measurements. We list the bar length measurements in Table~\ref{A3}.

In Figure~\ref{bar_length} panel a) and b), we present the star formation rate (SFR) vs. stellar mass plot for all barred galaxies in the MaNGA sample with the star formation main sequence relation of \citet{Davies16} for W3 in black. For reference, all galaxies in MPL-8 are also plotted as grey points. In panel a), data points are colour-coded by the length of the bar in kiloparsec. The correlation between bar size and galaxy stellar mass at low-redshift is apparent, as we see the most massive galaxies possess bars of longer physical length. This same trend has been shown from work with the Spitzer Survey of Stellar Structure in Galaxies \citep[S$^{4}$G;][]{Sheth10} sample of barred galaxies \citep{Diaz-Garcia16a,Erwin18}.

If instead, we divide the bar length by the effective radius of the galaxy, $R_{\rm{e}}$, we obtain a scale-free measurement of the fraction of the galaxy dominated by the bar. In panel b), the points are coloured by the bar length in units of $R/R_{\rm{e}}$. We note that some previous literature use the isophotal radius or disk scale-length to characterise the fractional size of bars within their host galaxies. We chose not to employ the disk scale-length, as it is inherently difficult to measure in barred galaxies, as the non-axisymmetric nature of the bar light profile tends to bias bulge and disk measurements. We did however, perform the same analysis as above using disk scale-length measurements from the \citet{Simard11} catalogue, and found similar results. 

From Figure~\ref{bar_length}, we see that for a given stellar mass, galaxies hosting fractionally long bars can be either star-forming or passive, while short bars are mainly hosted by systems that lie along the main sequence line. This trend is better shown in panel c) of Figure~\ref{bar_length}, where we plot the distance of a galaxy from the main sequence line as a function of fractional bar length, with points coloured by galaxy stellar mass. For a given fractional bar length, it is the more massive galaxies that are further from the main sequence line. We see little trend in fractional bar length with stellar mass.


\citet{Erwin19} discusses bar length correlations in detail and present a bi-modal scenario in which bar length is almost independent of stellar mass for low-mass galaxies ($\log( {\rm{M}_{\star}/\rm{M}_{\odot}})<10.1$), but correlates well for higher stellar masses. They also find that disk scale length and galaxy half-light radius correlate better with bar length than galaxy stellar mass. 
\citet{Diaz-Garcia16a} also report a bi-modal trend, confirming that fractional bar length is correlated with stellar mass for $\rm{M}_{\star}>10^{10}~\rm{M}_{\odot}$ but report an anti-correlation for $\rm{M}_{\star} < 10^{10}~\rm{M}_{\odot}$. We see no such trend in this work, though note that given disparate data sets and scaling measurements are discussed, we cannot directly compare.

This bi-modality in bar properties based on stellar mass presented in the literature paints a picture of two separate populations, the physics of which are determined by the stellar mass of the host galaxy \citep{Erwin18, Diaz-Garcia16b}.
Previous literature reports that bar properties are also strongly correlated with galaxy morphology \citep[e.g.][]{Diaz-Garcia16a}, such that S0s and early-type spirals host bars with longer scale lengths than late-type disk galaxies. Given the subjectivity of morphological classification, we chose not to follow this route, but note that given trends with stellar mass and distance from the star formation main sequence (SFMS) line, the galaxies in panel b) of Figure~\ref{bar_length} with the longest scaled bars will likely be the more passive S0s referenced by this literature. 

Regardless of morphology, the question remains of why the scaled length of bars in star-forming galaxies are shorter than their more passive counterparts. The reason may be that these bars have simply formed more recently, or took longer to grow.
\citet{Athanassoula13} argue that bars in gas-poor galaxies were in place 7-8 Gyr ago \citep[also][]{Sheth08, Kraljic12}, but gas-rich galaxies take longer to form bars, as recently as 4-5 Gyr ago. If we assume that bars grow in length with time \citep[e.g.][]{Elmegreen07,Gadotti11}, and that bars are not easily destroyed \citep[e.g.][]{Berentzen07, Villa-Vargas10}, then if gas is not accreted/replenished, the longest bars should be observed in the most gas-poor galaxies. 
We know from observations of higher-redshift low-mass spirals that the bar fraction is much lower at $z=0.5$ \citep{Abraham99} and $z=0.8$ \citep{Sheth08} than it is today. In higher mass galaxies, this is not the case however. This suggests that at $z\sim0.5-0.8$, the bars of high-mass galaxies were already in place, but for low-mass galaxies, bar formation is a more recent phenomenon \citep[e.g.][]{Kraljic12, Melvin14}. 

This result is also consistent with the work of \citet{Kruk18}, who use the bar length measurements of \citet{Hoyle11} to show that while bar effective radius increases with galaxy mass, so too does disk effective radius. So while bar physical length increases with mass, it is not well correlated with scaled bar length. This is also confirmed by observational results such as \citet{Sanchez-Janssen13}, \citet{Diaz-Garcia16b}, and \citet{Erwin19}, who show that the disks of barred galaxies are generally more extended than for non-barred counterparts. 

Although further investigation is required to draw a definitive conclusion, it would seem that as a galaxy grows in size, so too does its bar, but not more so than other components of a galaxy. Bars in more passive galaxies may have formed longer ago than those in more gas-rich galaxies, and this may be why they are longer. 

\begin{figure*}
\centering
\begin{subfigure}{0.49\textwidth}
\includegraphics[width=\textwidth]{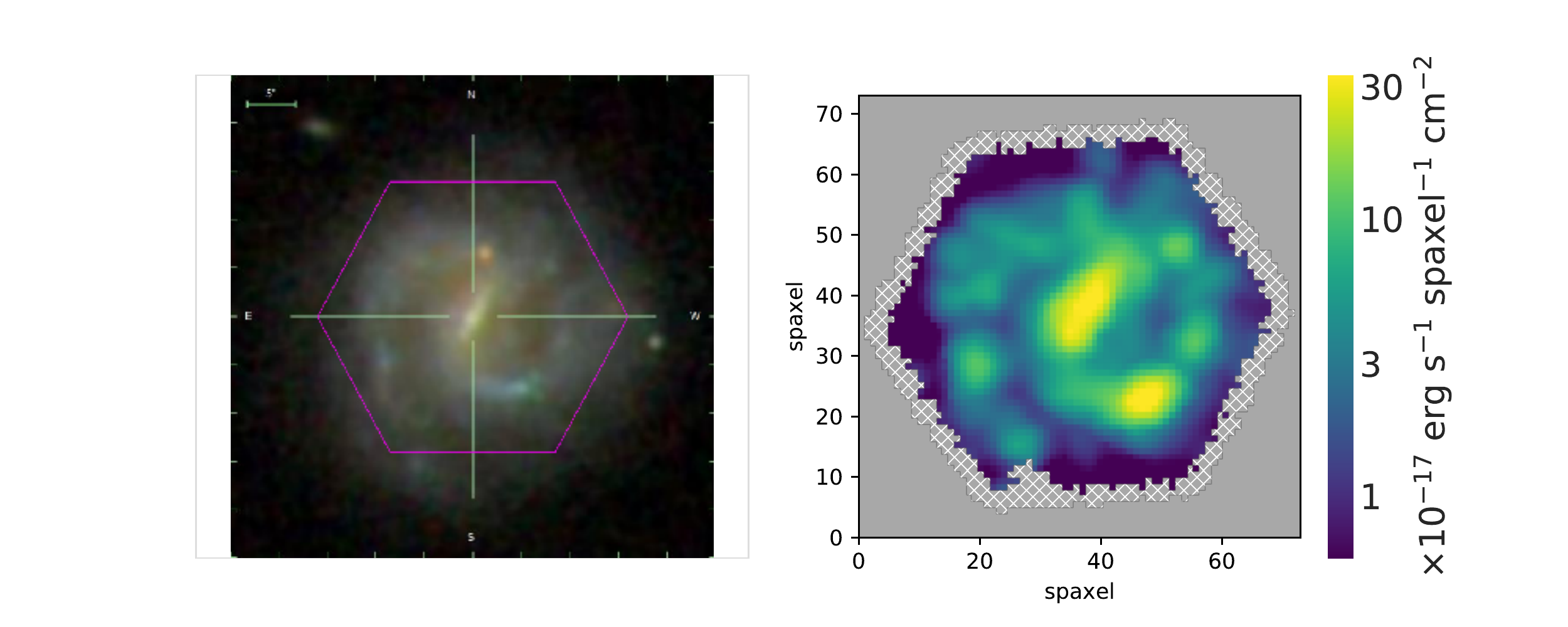} 
\caption{H$\alpha$ present along the bar - MaNGA galaxy 8465-12701.}
\end{subfigure}
\begin{subfigure}{0.49\textwidth}
\includegraphics[width=\textwidth]{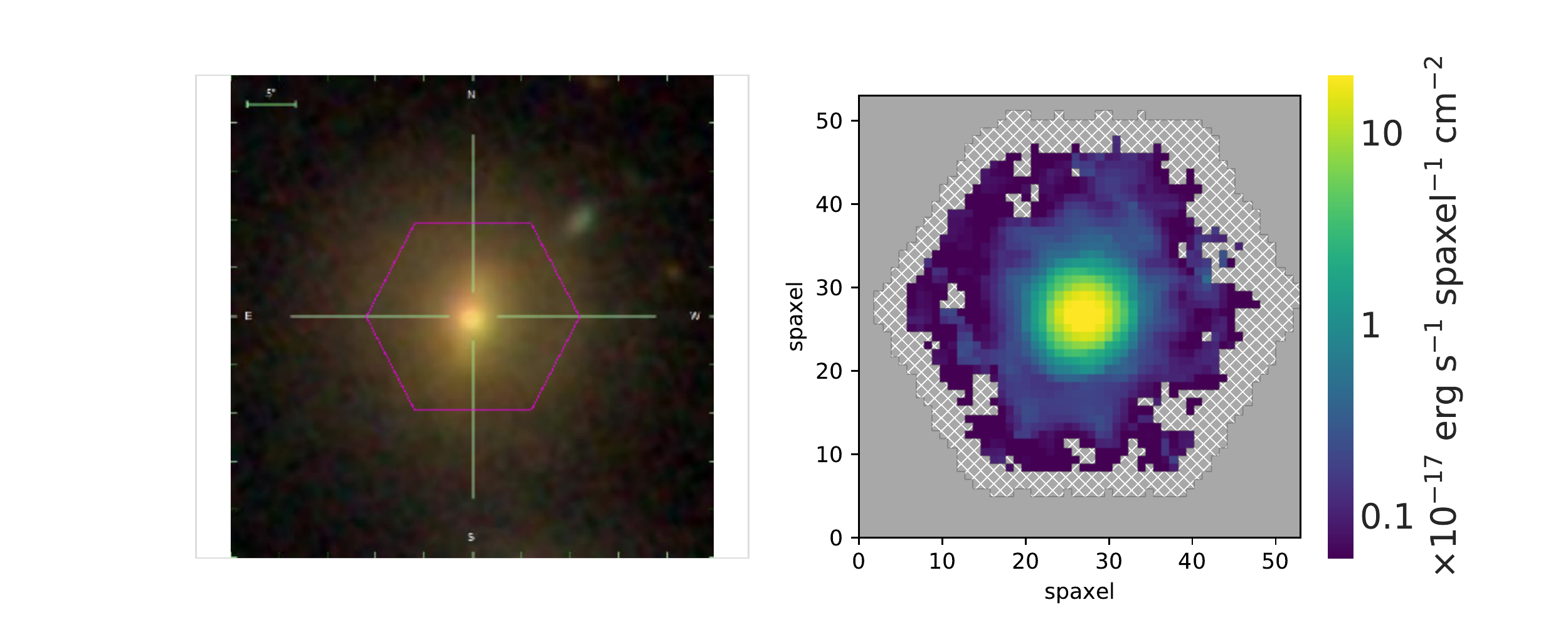} %
\caption{Predominantly central H$\alpha$ - MaNGA galaxy 8595-6104.}
\end{subfigure}
\hfill
\begin{subfigure}{0.49\textwidth}
\includegraphics[width=\textwidth]{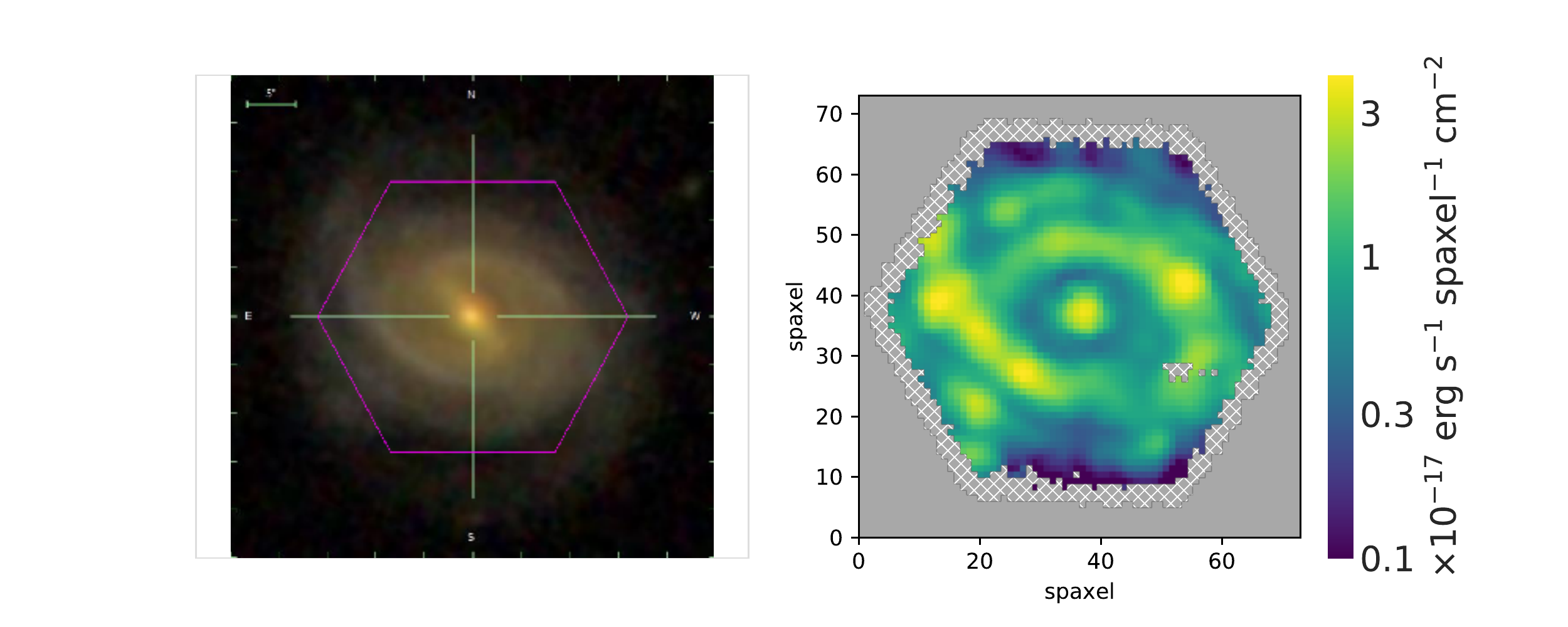} %
\caption{Prominent ring - MaNGA galaxy 8452-12703.}
\end{subfigure}
\begin{subfigure}{0.49\textwidth}
\includegraphics[width=\textwidth]{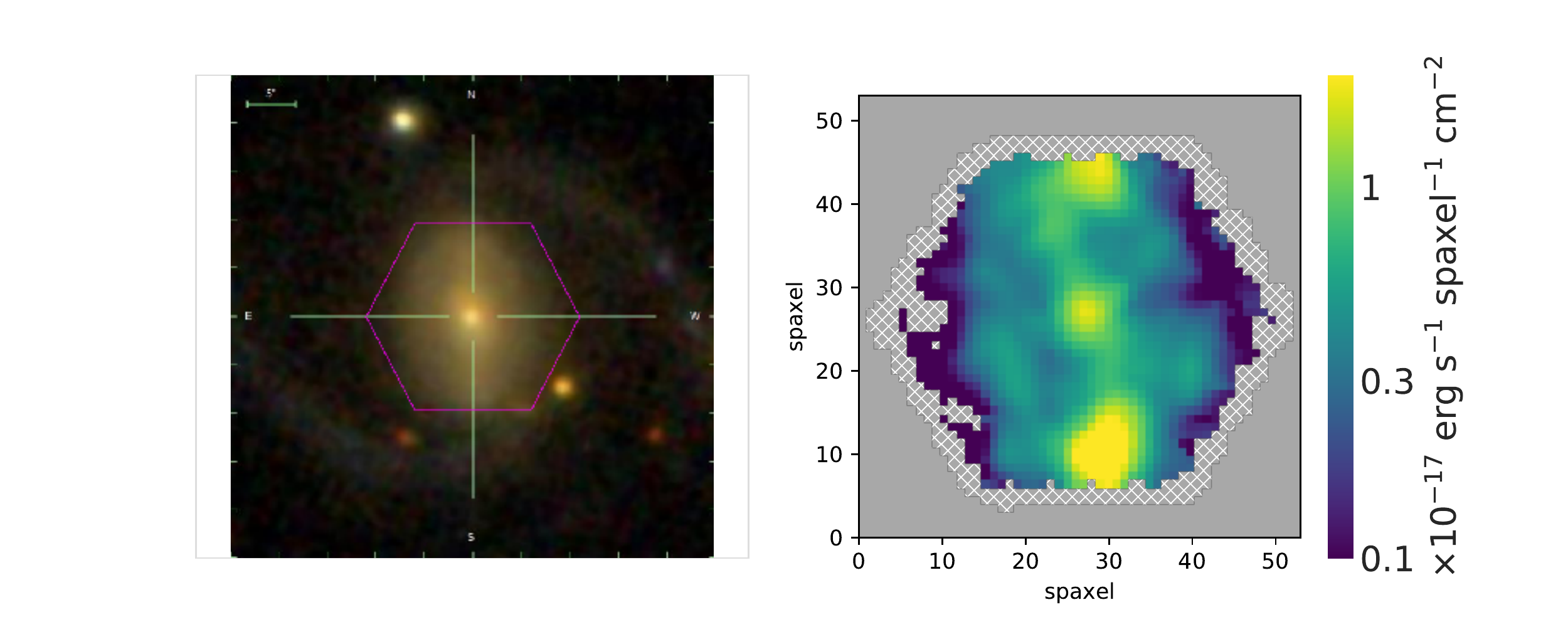} %
\caption{H$\alpha$ at ends of the bar (and sometimes the centre) - MaNGA galaxy 8135-6103.}
\end{subfigure}
\hfill
\begin{subfigure}{0.49\textwidth}
\includegraphics[width=\textwidth]{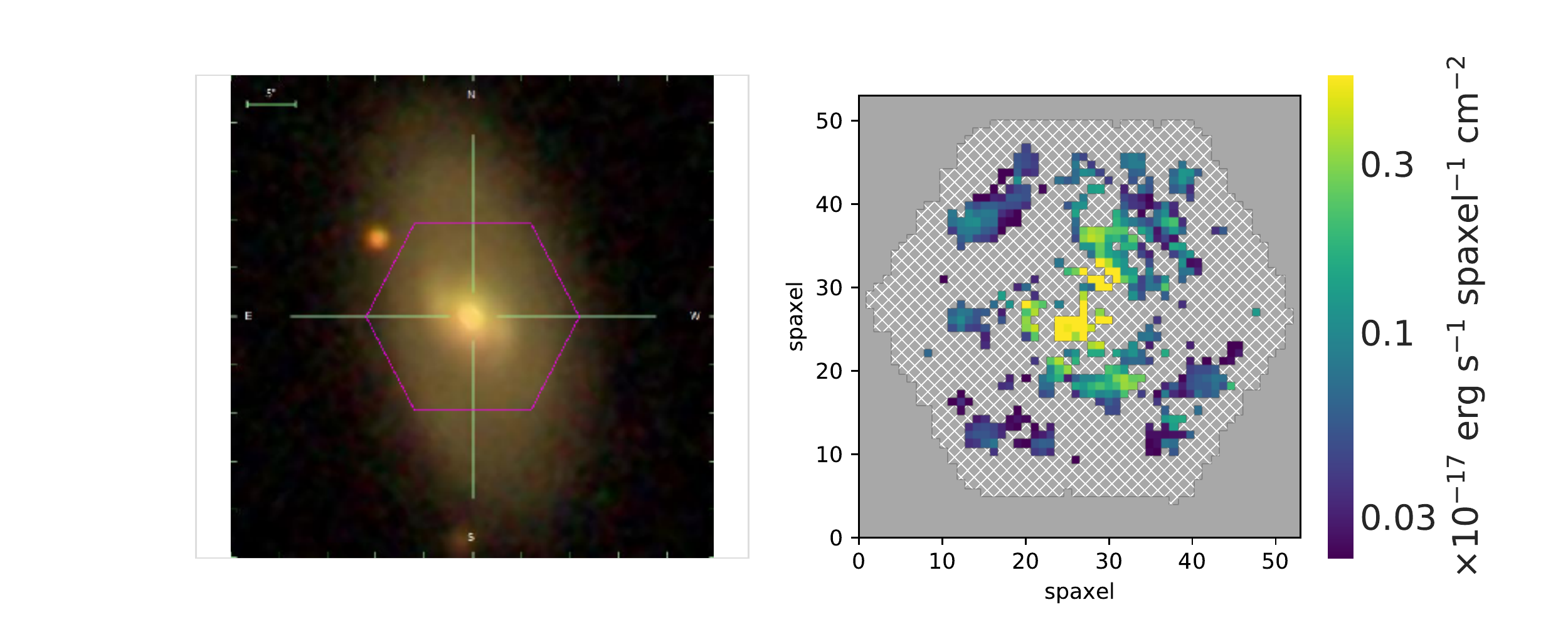} %
\caption{No significant H$\alpha$ - MaNGA galaxy 8243-6103.}
\end{subfigure}
\begin{subfigure}{0.49\textwidth}
\includegraphics[width=\textwidth]{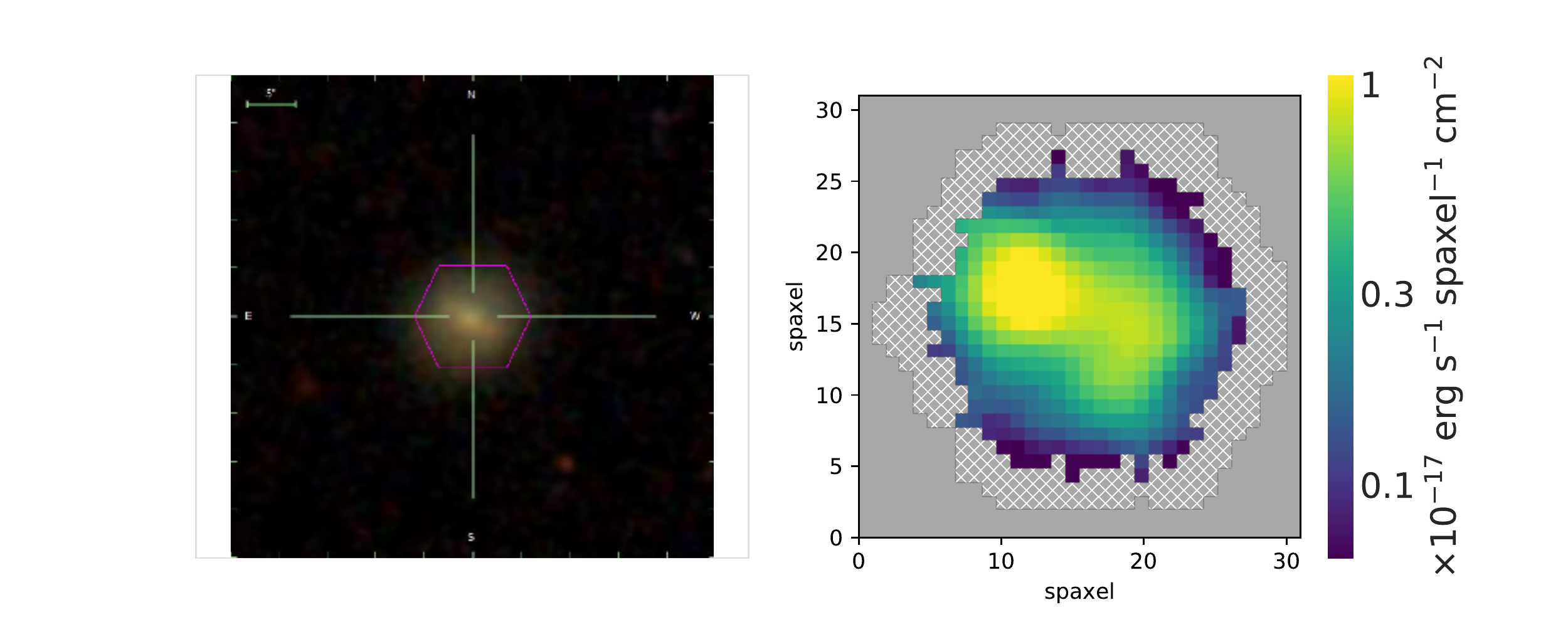} %
\caption{Unclassifiable - MaNGA galaxy 8992-1902.}
\end{subfigure}

\caption{Examples of the six H$\alpha$ morphology classifications devised for this work, with galaxies denoted by their MaNGA plate and IFU. For each class, we show the SDSS $gri$ image of the galaxy with MaNGA field of view overlaid in magenta, and the logarithmically-scaled H$\alpha$ flux map.} %
\label{Ha_morph_egs}
\end{figure*}

\section{Classifying Ionized Gas Morphology}
\label{classifications}
From their position relative to the SFMS line in Figure~\ref{bar_length}, we know that the majority of MaNGA barred galaxies are star-forming, and should hence contain quantities of ionised gas. Given the spatial resolution of MaNGA, we are able to investigate the position and morphology of ionised gas for a large sample of barred galaxies. This is made possible by employing the H$\alpha$ emission line maps provided by the MaNGA DAP, logarithmically scaled, created by measuring the Gaussian profile integrated flux from a combined continuum + emission line fit \citep[for details, see][]{Westfall19, Belfiore19}. 

Motivated by works such as \citet{Verley07} and \citet{Neumann19}, we devise a visual classification scheme, whereby the H$\alpha$ map of a barred galaxy falls into one of six categories. Example maps for each category are shown in Figure~\ref{Ha_morph_egs}, and described below, with the percentage that each category contributes to the overall galaxy sample and binomial errors also listed:
\begin{itemize}
\item \textbf{H$\alpha$ present along bar (panel a; 18 $\pm$ 1\% of sample).} Extended H$\alpha$ emission was detected across the galaxy, coinciding with (or close to) the position of the bar in the optical $gri$ image of the galaxy. If any H$\alpha$ emission was seen along the bar (even if also present in other regions), it was classified into this category.

\item \textbf{Predominantly central emission (panel b; 20 $\pm$ 2\% of sample).} Resolved H$\alpha$ is concentrated chiefly in the central regions of the galaxy. 

\item \textbf{Prominent ring (panel c; 21 $\pm$ 2\% of sample).} The galaxy possesses a prominent ring of H$\alpha$ emission, generally coincident with a ring in the optical image, and frequently with H$\alpha$ emission in the central region also. The emission around the ring is mostly uniform, and there is no over-concentration of H$\alpha$ at the ends of the bar

\item \textbf{H$\alpha$ at the ends of the bar, or centre and ends (panel d; 18 $\pm$ 1\% of sample).} There is significant H$\alpha$ emission at the ends of the bar, and more commonly, the central region and the ends of the bar, but not along the rest of its length, as seen by previous studies such as \citet{Verley07}. We also note that other features such as faint rings may be present, but this category is characterised by an overdensity of H$\alpha$ at the ends of the bar. 
 
\item \textbf{No H$\alpha$ detected (panel e; 10 $\pm$ 1\% of sample).} No significant H$\alpha$ emission is detected in this galaxy.

\item \textbf{Unclassifiable (panel f; 13 $\pm$ 1\% of sample).} The H$\alpha$ emission could not be classified into any of these categories, usually because the spatial resolution was not sufficient for morphological classification, or there were multiple (sometimes merging) galaxies in the IFU.

\end{itemize}

\begin{figure*}
\centering
\begin{subfigure}{0.99\textwidth}
\includegraphics[width=\textwidth]{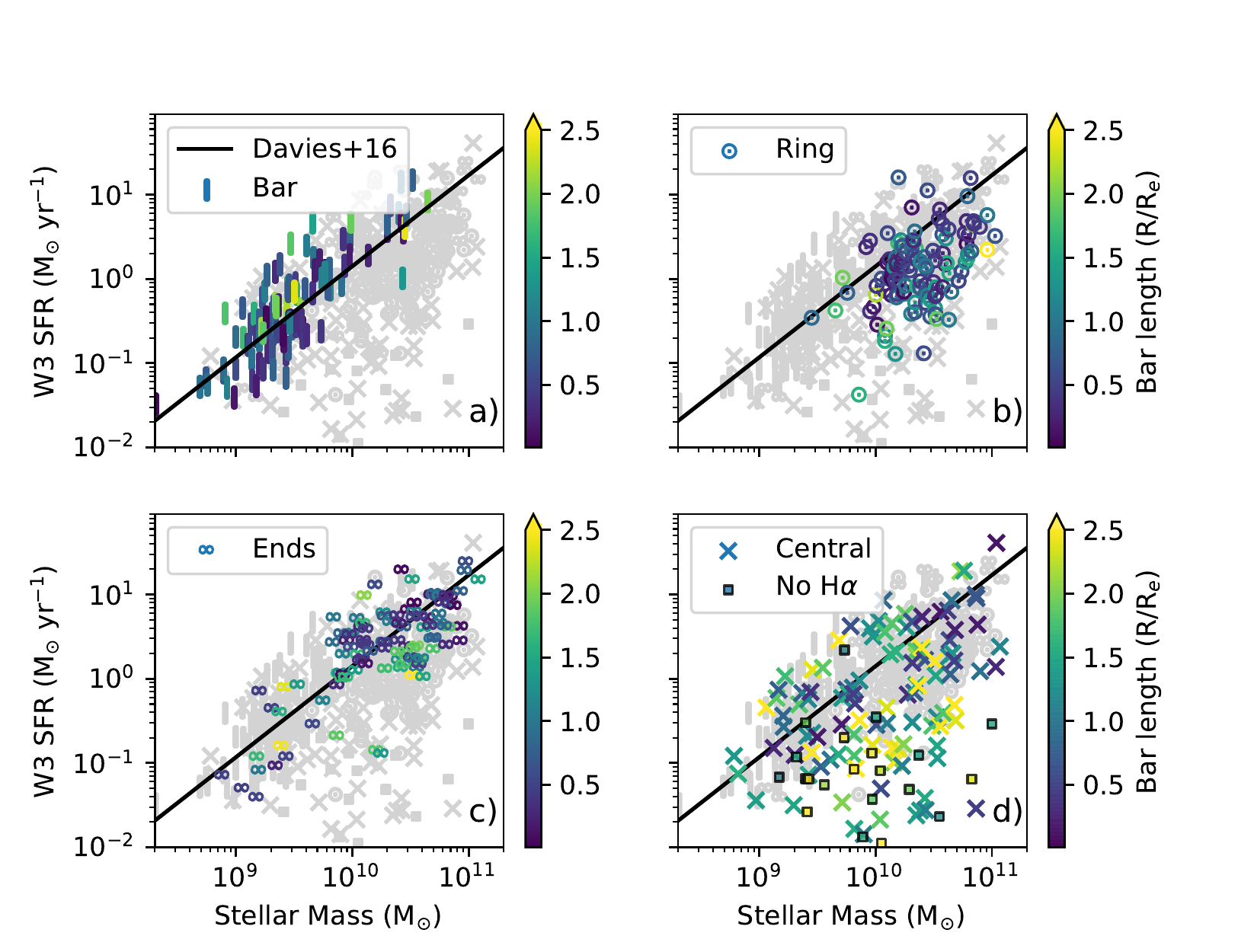}  
\end{subfigure}
\caption{SFR-$\rm{M}_{\star}$ diagram for MaNGA barred galaxies. Each panel highlights a different H$\alpha$ morphology, and points are coloured by their scaled bar length. The black line is the W3 star formation main sequence (SFMS) relation from \citet{Davies16}.
Galaxies with barred H$\alpha$ morphology are lower-mass, and lie on (or sometimes slightly above) the SFMS line. Ringed H$\alpha$ morphology belong to higher-mass galaxies that lie below the SFMS line. Galaxies with H$\alpha$ at the ends of their bars are also generally higher-mass, and lie close to the main sequence line. Central H$\alpha$ galaxies are distributed across the SFR-stellar mass plane, and galaxies with no H$\alpha$ lie mostly below the main sequence line. Galaxies with central or no H$\alpha$ possess bars of longer scale length than the other H$\alpha$ morphologies.}
\label{Ha_morph_CMD}
\end{figure*}

Each galaxy was classified by one author (AFM) and the features in the H$\alpha$ maps noted and given a numeric value, described in Table~\ref{A1}. When a galaxy possessed more than one feature, a hierarchy was developed such that the presence of H$\alpha$ along the bar would automatically place a galaxy into category a): H$\alpha$ present along a bar. If an overdensity of H$\alpha$ was seen at the ends of the bar (but not along the bar), it was always placed in category d): H$\alpha$ at the ends of the bar. Similarly, if a ring feature was seen (but no H$\alpha$ along the bar nor concentrated at the ends) it was classified into category c): prominent ring. If no other of the above-mentioned features were detected but central H$\alpha$ was present, a galaxy was classified into category b): predominantly central emission. Table~\ref{A2} shows which H$\alpha$ category a given combination of numeric values described in Table~\ref{A1} were placed into. We investigated any possible trends in classification with IFU bundle size and found no biases towards any bundle size for a given H$\alpha$ morphology. Most importantly, category f): Unclassifiable, was not biased towards small IFU bundle sizes. Additionally, we note that we do not expect the radius of a bar to extend beyond MaNGA's field of view. Indeed, when we compared the scaled bar lengths to the expected IFU coverage of each galaxy (1.5 $R_{e}$ for the Primary+ sample and 2.5 $R_{e}$ for the Secondary sample), we found only nine occasions where the bar length was of order the IFU size. The H$\alpha$ morphology of each galaxy is listed in Table~\ref{A3}.

We investigate the correlation between galaxy H$\alpha$ morphology and position on the SFR-$\rm{M}_{\star}$ diagram in Figure~\ref{Ha_morph_CMD}. The shape of each point in this Figure indicates the H$\alpha$ morphology of a galaxy, and points are coloured by the scaled length of the bar. Each panel highlights a certain H$\alpha$ morphology, with the other morphologies plotted in grey, to show the overall distribution. 

In general, galaxies with H$\alpha$ along their bar are found along (or slightly above) the SFMS, and are mostly lower-mass galaxies (94\% with $\rm{M}_{\star}<10^{10}~\rm{M}_{\odot}$). Galaxies with a ring of H$\alpha$ tend to lie slightly below the SFMS line, and consist of mostly higher-mass galaxies (91\% with $\rm{M}_{\star}>10^{10}~\rm{M}_{\odot}$). When H$\alpha$ is present at the ends of a bar, galaxies are mostly on the main sequence line, and 74\% have $\rm{M}_{\star}>10^{10}~\rm{M}_{\odot}$. In contrast, galaxies with central H$\alpha$ are found all over SFR-$\rm{M}_{\star}$ parameter space, possibly because of the likelihood of distinct ionisation mechanisms for the observed gas, including star formation, active galactic nuclei (AGN), or low ionisation nuclear emission line region (LINER) activity. Finally, galaxies with no discernible H$\alpha$ are found almost exclusively below the SFMS line, and at all stellar masses. 

\begin{figure}
\centering
\begin{subfigure}{0.49\textwidth}
\includegraphics[width=\textwidth]{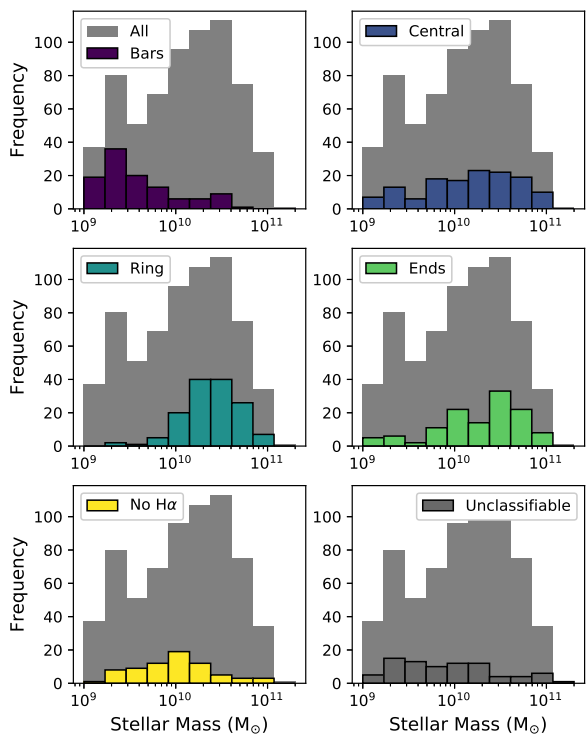}  
\end{subfigure}
\caption{Histograms of the stellar mass distributions of the six categories of H$\alpha$ morphology. Each panel highlights a different H$\alpha$ morphology, and the entire sample is shown behind in grey. Low-mass galaxies are more likely to possess H$\alpha$ along their bars, while high-mass galaxies are likely to host H$\alpha$ at the ends of their bars, or in ring-shaped morphologies.}
\label{Ha_morph_stell_mass_hist}
\end{figure}

Trends with mass are better seen in Figure~\ref{Ha_morph_stell_mass_hist}, which shows histograms of the galaxy stellar mass distribution for each H$\alpha$ morphology category. The distribution of galaxies with H$\alpha$ along the bar is heavily skewed towards low-mass galaxies, while for galaxies with H$\alpha$ rings and H$\alpha$ at the ends of the bar, the opposite is true. This dichotomy in mass distribution is significant between H$\alpha$ morphologies, and we attempt to explain these results given a bar evolution scenario below.

\subsection{Insights into Bar Evolution from H$\alpha$ Morphology}

Observations of central H$\alpha$ can be explained by central starbursts (or the remnants of), fuelled by gas deposition onto the centre of a galaxy along bars. This has often been reported in the literature 
\citep[e.g.][]{Devereux87, Telesco93, Knapen95, Alonso-Herrero01, Wang12, Florido15, Lin17, Chown19}. In general, galaxies with central or no H$\alpha$ possess bars that extend over a larger fraction of the overall galaxy. This could be because bars in more passive galaxies formed earlier, grew more quickly, or that larger bars are more effective at quenching their host galaxy.

H$\alpha$ concentrations at the ends of the bar are common in high-mass galaxies, and these galaxies lie mostly on the SFMS, indicating that they are still forming stars. H$\alpha$ (and presumably, star formation) concentrations at the ends of a bar have been shown in simulations to be due to a combination of kpc-scale dynamics (gas flows, shear),  and parsec-scale turbulence, along with cloud collisions \citep{Renaud15}. For star formation to occur in a galaxy, several physical properties must be present, including the presence of cold, dense gas. Equally, certain physical conditions must be absent, and one of these is shear motions in the ISM. 

Simulations show that the elongated orbital motions of the stars within a bar induce shear in the ISM \citep{Athanassoula92, Emsellem15}. Shear prevents the formation of molecular clouds, despite the presence of dense gas. At the ends (and sometimes edges) of a bar, however, \citet{Renaud15} show in a hydrodynamical simulation of a Milky Way-mass barred galaxy that weaker shear balances with supersonic turbulence (evidenced by high Mach numbers in these regions), to allow dense gas to condense further into molecular clouds. 
The slowdown in orbital velocity at the ends of the bar as stars on x1 orbits turn back toward the galactic centre allows for the survival of these structures until the cloud fragmentation can take place. It seems that in high-mass galaxies at least, the presence of star formation at the ends of a bar (or at the ends and central regions) can be well explained by a theoretical framework. Indeed, these structures have already been observed and reported in the literature for small samples of galaxies \citep[e.g.][]{Reynaud98,Verley07,Neumann19}.

The ring-like features in the H$\alpha$ maps are also present in galaxies with stellar mass $>10^{10}\rm{M}_{\odot}$, lying both on and below the SFMS, and always coincident with the presence of a ring in the optical image of the galaxy. This stellar mass constraint is perhaps not surprising, given inner rings are generally mostly found in massive barred galaxies \citep[e.g.][]{Herrera-Endoqui15, Diaz-Garcia19}. The interplay between inner rings and bars is complex: while they commonly occur together, they are both also found unaccompanied in galaxies \citep[e.g.][]{Comeron14}. This, and the fact that \citet{Grouchy10} found little difference between the star formation activity of inner rings in barred and unbarred galaxies points towards a scenario in which there is no causal relation between star formation in rings and the presence of a bar. 
When the opposite scenario is considered however, observations have shown a correlation between the lack of star formation in a bar and the presence of an inner ring \citep{Neumann19}.
From this we infer that some gas and stars have been trapped in resonance rings, most likely as the result of the redistributive effects of a strong bar \citep{Schwarz81}. This trapped gas must be maintained under the correct conditions to form stars, and is thus prevented from flowing along the bar into the central galactic regions.

H$\alpha$ along a bar is only seen in low-mass galaxies of $\rm{M}_{\star}\lesssim10^{10}\rm{M}_{\odot}$. Few simulations have been made of such lower-mass systems \citep[e.g.][]{Carles16}, but one would expect that the shear motions within the bars are not strong enough to prevent the formation and collapse of molecular clouds to form stars. This is summarised in \citet{Jogee02}, who hypothesise that weaker shocks and shear can induce star formation at the leading edges of bars, rather than inhibit it.
 Observationally, \citet{Seigar05} find a connection between star formation rate suppression and shear within spiral arms from infrared observations of 33 spiral galaxies. They derive a critical shear rate, above which star formation turns off in the disks of spiral galaxies. A similar behaviour might be expected within the bars of these galaxies. 

\subsection{Comparison to Previous Literature}
A previous attempt to understand H$\alpha$ morphology in terms of an evolutionary sequence was made by \citet{Verley07}, who classified the H$\alpha$ distributions of a sample of 45 barred galaxies. They defined three main categories, the first of these classes was designated Group E, and contained galaxies with bright, central H$\alpha$ and H$\alpha$ in knots at the ends of the bar and in the spiral arms. The second category are galaxies that show no central H$\alpha$ emission, and very little gas elsewhere, designated as Group F. Finally, the third group consists of galaxies containing H$\alpha$ emission along the bar, designated Group G. They present a bar evolution argument whereby as a bar forms from a disk instability from inflowing gas, the gravitational torque within the bar drives gas inflow towards the centre (Group G bars). The next phase is a transition from G to E phase, where they mention a ring can form, and finally, a galaxy ends up as Group F, as all gas is depopulated from the bar. \citet{Verley07} go on to predict the gas infall destroys the bar. This scenario is supported by observations of a higher fraction of both bars and lenses in S0 galaxies compared to spirals \citep{Laurikainen13}.

However, the mass dependence of H$\alpha$ morphology presented here makes the evolutionary sequence proposed by \citet{Verley07} unlikely, as the short, star-forming bars embedded in low-mass galaxies will not evolve into higher-mass disks without tidal disruptions destroying any resonance features present. That said, for high-mass galaxies, a scenario could be imagined in which more and more gas gets accumulated into a ring feature, forms stars, and at the same time, gas is consumed globally in the galaxy such that it moves away from the SFMS line. This gas consumption may or may not be accelerated by the presence of a bar, but the result may be that these galaxies end up below the main sequence line once their star formation has ceased (with `no H$\alpha$' morphologies).

What is apparent is that there is a complex interplay between galaxy stellar mass, bar length, H$\alpha$ morphology, and SFR. In addition, the effects of HII gas fraction and depletion time \citep{Saintonge12} and central mass concentration prominence \citep[e.g.][]{Bell12} on the overall galaxy quiescence have not been considered here. While these factors may also be involved in the trends seen in Figure~\ref{Ha_morph_CMD}, we can still conclude that stellar mass and H$\alpha$ morphology seem to have a strong connection, and it is the most passive galaxies with central or no H$\alpha$ that host the longest bars.

\begin{figure*}
\centering
\includegraphics[width=0.9\textwidth]{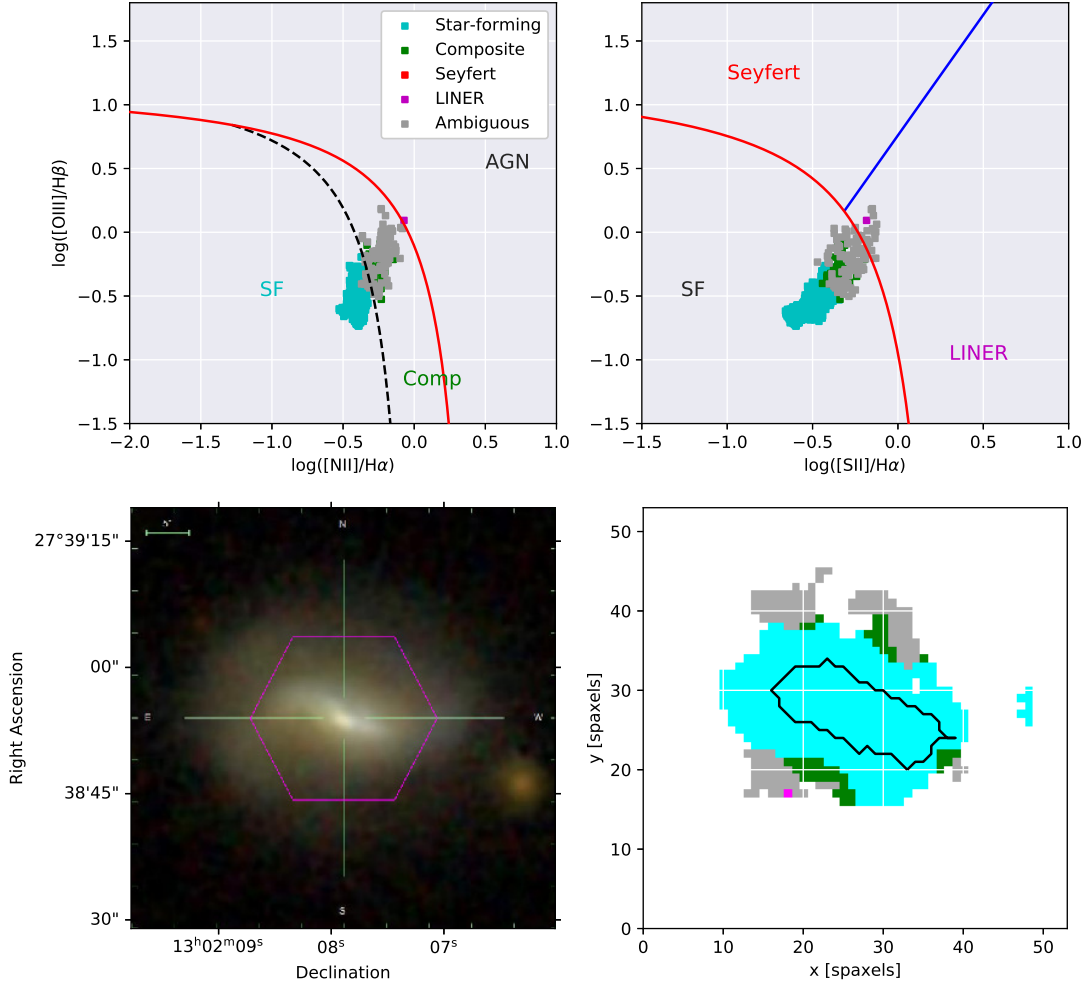} 
\caption{Example BPT diagrams and map for MaNGA galaxy 8935-6104. On the top row are [NII] (left) and [SII] (right) BPT diagrams of every spaxel above a signal to noise cut in the IFU field of view. Spaxels are colour-coded by the region that the lie on the BPT diagram. Below is the SDSS $gri$ image of the galaxy with MaNGA field of view overlaid in magenta (left) and [NII] BPT map of the galaxy with a GZ:3D bar mask overlaid in black, with spaxels coloured by the ionisation mechanism as in the BPT diagrams. As all of the spaxels within the GZ:3D bar mask are cyan, the dominant ionisation mechanism in the bar is star formation.}
\label{BPT_example}
\end{figure*}

\section{Systems with H$\alpha$ Along the Bar}
The galaxies with H$\alpha$ observed along the length of the bar provide a fascinating subsample. Observations of neutral gas, CO, and ionised gas along bars for small samples of galaxies exist \citep[e.g.][]{deVauc63,Regan99,Sorai00,Popping10}, implying that gas flow occurs along bars in line with a bar quenching scenario. The presence of ionised gas however is interesting, and the source of this ionisation is equally intriguing. For this reason, we investigate the ionisation mechanism of gas along the bars of this subsample, comprised of 115 galaxies. \\

\subsection{Ionisation Mechanism}
\label{ionization_sect}

H$\alpha$ is mainly produced in the \textsc{HII} regions surrounding O- and B-type stars, although shocks and AGN or LI(N)ER activity can also contribute to observed H$\alpha$ emission.
In order to determine the dominant ionisation mechanism of the gas within the bars, we make use of Galaxy Zoo: 3D (GZ:3D; Masters et al. \textit{in prep}) data. GZ:3D is a citizen science project that asks participants to draw regions around various galaxy components they see on SDSS galaxy images, which are then translated into masks that may be applied directly to MaNGA data cubes. These masks can be used to separate spaxels from a galaxy data cube whose light is dominated by individual components including bulges, bars, spiral arms, and foreground stars. 

For each galaxy, we determine the ionisation mechanism for every bar spaxel according to a Baldwin, Phillips and Terlevich \citep[BPT;][]{Baldwin81} classification. We apply the bar masks to the BPT diagrams, and to eliminate any poor classifications, take only spaxels which $>80\%$ of respondents have deemed to be within the bar region of a galaxy. The mechanism with the highest fraction is deemed the dominant ionisation mechanism for the gas in the bar.

\begin{figure*}
\centering
\begin{subfigure}{0.99\textwidth}
\includegraphics[width=\textwidth]{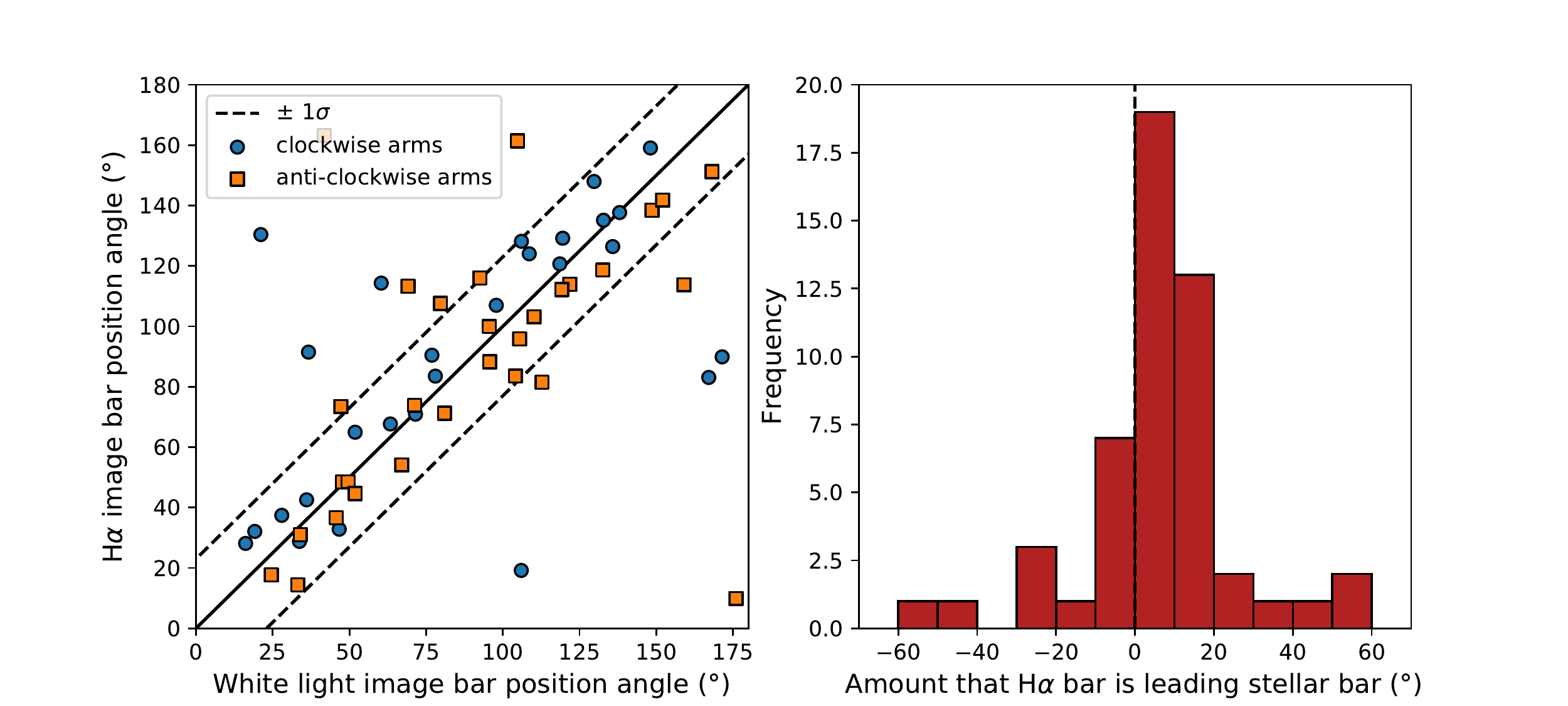}  
\end{subfigure}
\caption{Differences between H$\alpha$ and stellar bar position angles. On the left, we plot the H$\alpha$ image bar position angle against the white light bar position angle, with points colour-coded by the direction of rotation of the galaxy. A solid black 1:1 line is shown, along with dashed 1$\sigma$ lines. Galaxies that rotate clockwise have the H$\alpha$ bar ahead of the stellar bar if they are above the black 1:1 line, and the opposite is true for anticlockwise-rotating galaxies. On the right is a histogram of the amount by which the H$\alpha$ bar is leading the stellar bar (negative values indicate it is trailing). The majority of H$\alpha$ bars lead their stellar bars by 0--20\textdegree.}
\label{bar_differences}
\end{figure*}

Emission line maps are accessed through \textsc{Marvin}, a software tool specifically designed to visualise MaNGA data \citep{Cherinka18}. \textsc{Marvin} has the ability to create BPT diagrams on a spaxel-by-spaxel basis, then display a map of a particular galaxy with the ionisation mechanism of each spaxel shown. \textsc{Marvin} uses the \citet{Kewley06} star-forming region in the [SII] and [OI] BPTs, and the \citet{Kauffmann03} star-forming region for the [NII] BPTs. 
The [OI] line is weak compared to [NII] and [SII] in most galaxy spectra, and often cannot be reliably measured. For this reason, we chose not to employ the [OI] BPT diagram. Hence, a spaxel is only classified if it meets the criteria for a particular classification in both the [SII] and [NII] BPTs. We refer the reader to \citet{Cherinka18} for a more thorough explanation of \textsc{Marvin}'s capabilities, and to \citet{Belfiore19} for a discussion of error modelling in MaNGA emission line measurements.
An example output from \textsc{Marvin} BPT mapping function is shown in Figure~\ref{BPT_example} for MaNGA galaxy 8935-6104. The top panels show the [NII] and [SII] BPT diagrams for every spaxel above the signal-to-noise threshold for this galaxy with spaxels colour-coded by the region of the BPT diagram that they lie in. The bottom left panel shows an optical image of the galaxy 8935-6104, with the MaNGA hexagonal-shaped field of view shown in magenta. The bottom right panel shows the BPT map for this galaxy, with spaxels colour-coded by their BPT classification and GZ:3D bar mask overlaid. 

Of the 115 galaxies with H$\alpha$ present along the bar, 109 have bars dominated by star formation. The remaining six are dominated by composite regions.
We conclude the dominant ionisation mechanism of gas within bars is star formation.
This result agrees with previous studies of spatially resolved star formation in barred galaxies, showing that the SFR of barred galaxies may be enhanced in the centre by the presence of a bar. i.e. the H$\alpha$ emission is star-forming, not AGN or LINER-dominated \citep{Catalan-Torrecilla17}.

\subsection{Spatially-resolved Star Formation Along Bars}
\label{bar_angles}
We have confirmed that star formation is occurring along the bars of a select number of MaNGA barred galaxies. 
Previous literature finds tentative evidence that the \textsc{HII} regions of barred galaxies are often offset towards the leading edge of a bar \citep[usually by between 5\textdegree--30\textdegree~e.g.][]{Martin97, Sheth02, Verley07, Neumann19}. 
However, these studies consisted of only eleven, six, 27, and eight late-type disk galaxies, respectively. Sometimes the opposite has been reported -- for example, \citet{Zurita08} imply that massive stars form on the trailing edge of a bar dust lane and instead migrate towards the leading edge over time. 
Building on this previous work, we measure the differences in bar position angle between the H$\alpha$ image and the collapsed MaNGA data cube, or `white light' image for the 115 galaxies with ionised gas visible along their bars. Given that the white light image should be dominated by old stellar populations while the H$\alpha$ image traces new stars, any offset in bar position angle between the two should indicate preferential locations of star formation along the bar.

The same Fourier decomposition code modified from that of \citet{Kraljic12} that was used to measure bar lengths also returns the position angle of the bar in an image. We measure the bar position angle relative to East in the white light and H$\alpha$ images and determine the difference between the two. Stellar and H$\alpha$ bar length and position angle measurements are listed in Table~\ref{A3}. In some cases, either the stellar (three cases) or H$\alpha$ (four cases) bar could not be measured by the Fourier analysis code, typically because the light was too clumpy, or the code picked up H$\alpha$ in the spiral arms in addition to the bar and could not find a suitable cutoff point. 
From visual inspection of SDSS $gri$ images, we determine the direction of galaxy rotation in the line of sight, and designate them as either clockwise, anticlockwise, or unknown. We assume here that spiral arms trail as a galaxy rotates. Lenticular galaxies, and galaxies with prominent rings but no spiral arms, along with irregular spirals were all classed into the `unknown' category. 

In panel a) of Figure~\ref{bar_differences}, we present the stellar and H$\alpha$ bar position angles for galaxies in the H$\alpha$-barred subsample. Galaxies deemed to be rotating clockwise are shown in blue, and anticlockwise in orange. The solid black 1:1 line is shown, along with $1\sigma$ of the distribution either side. A galaxy rotating clockwise will have a H$\alpha$ bar ahead of its stellar bar if its point is above the 1:1 line, and anticlockwise-rotating galaxies will be below the 1:1 line if the H$\alpha$ is ahead. 

Panel b) of Figure~\ref{bar_differences} is a histogram of the angle by which the H$\alpha$ bar is leading the stellar bar. The angle is measured in the direction of rotation, so positive values indicate the H$\alpha$ bar is ahead of the stellar bar. We expect that the galaxies for which the H$\alpha$ bar is measured to be trailing the stellar bar are probably due to measurement scatter, given the width of the 1$\sigma$ error lines in panel a). Even with scatter, the majority ($67 \pm 7\%$) of H$\alpha$ bars measured lead their stellar bar by $0-20$\textdegree. 
 
Through observations of small numbers of galaxies, it has frequently been shown that HII regions are preferentially offset ahead of molecular gas and dust lanes in spiral arms \citep[e.g.][]{Vogel88,Rand92,Knapen96}. This is interpreted as star formation occurring at the leading edge of spiral arms, due to shocks in the compressed ISM. We expect a similar scenario within the bars of barred galaxies, where shear and turbulence forces prevent star formation from occurring anywhere but the leading edge of the bar \citep{Emsellem15, Renaud15}.

\section{Summary \& Conclusions}
We analyse the physical properties of 684 barred galaxies in the MaNGA galaxy survey, selected using Galaxy Zoo 2. We measure bar length in both MaNGA white-light and H$\alpha$ images and classify the H$\alpha$  maps into one of six categories: H$\alpha$ along the bar, predominantly central emission, a prominent ring of H$\alpha$, H$\alpha$ at the ends of the bar, no significant H$\alpha$, and unclassifiable H$\alpha$ maps. Our main findings are:
\begin{itemize}
\item Physical bar length correlates with galaxy mass such that higher-mass galaxies possess physically longer bars. When the bar length is normalised by the galaxy effective radius, we find the scaled bar length is better correlated with distance from the SFMS. More passive galaxies possess bars that occupy a larger fraction of their disks.

\item The six categories of H$\alpha$ morphology lie in distinct regions of the SFR-$\rm{M}_{\star}$ mass diagram, and we infer this is because different processes are occurring within these galaxies. 
\begin{itemize}
\item We do not see star formation along the bar of high-mass galaxies, and we speculate that this is due to the complex interplay between turbulence and shear in bar regions.

\item We witness star formation in rings of high-mass galaxies in $21 \pm 2 \%$ of the sample, and these are coincident with visual stellar rings in optical images. We suggest that gas is caught in resonances in these galaxies, and prevented from funnelling towards the centre.

\item H$\alpha$ emission is observed at the ends of the bar in $18 \pm 1 \%$ of the sample, consistent with simulations which show that this morphology can be due to both the buildup of gas as orbits turn around at the end of a bar, coupled with lower shear in these regions. 

\item Finally, for low-mass galaxies, star formation often occurs along the bar, and this star formation is generally found at the leading edge of the bar, consistent with a picture whereby gas is compressed, and then shocked into star formation.
\end{itemize}
We note that given the stringent Galaxy Zoo 2 selection criteria employed in this work, our result will likely be biased towards galaxies that host stronger bars. Although detailed simulations of gas-rich barred galaxies of a range of masses are required to confirm the physical processes at play, we already see that the properties of gas within low-mass barred galaxies is different from high-mass galaxies, due to the galaxy internal dynamics and strength of shear flows within bars. Mass seems to be a good indicator of the internal properties on barred galaxies, though bar length is better correlated with distance from the star formation main sequence line.

\end{itemize}

\section{Acknowledgements} 
The authors wish to thank the anonymous referee whose comments improved the quality of this work. The authors also thank Johan Knapen for helpful comments.

This research made use of Marvin, a core Python package and web framework for MaNGA data, developed by Brian Cherinka, Jos\'{e} S\'{a}nchez-Gallego, Brett Andrews, and Joel Brownstein. (MaNGA Collaboration, 2018).

Funding for the Sloan Digital Sky Survey IV has been provided by the Alfred P. Sloan Foundation, the U.S. Department of Energy Office of Science, and the Participating Institutions. SDSS-IV acknowledges
support and resources from the Center for High-Performance Computing at
the University of Utah. The SDSS web site is www.sdss.org.

SDSS-IV is managed by the Astrophysical Research Consortium for the 
Participating Institutions of the SDSS Collaboration including the 
Brazilian Participation Group, the Carnegie Institution for Science, 
Carnegie Mellon University, the Chilean Participation Group, the French Participation Group, Harvard-Smithsonian Center for Astrophysics, 
Instituto de Astrof\'isica de Canarias, The Johns Hopkins University, 
Kavli Institute for the Physics and Mathematics of the Universe (IPMU) / 
University of Tokyo, Lawrence Berkeley National Laboratory, 
Leibniz Institut f\"ur Astrophysik Potsdam (AIP),  
Max-Planck-Institut f\"ur Astronomie (MPIA Heidelberg), 
Max-Planck-Institut f\"ur Astrophysik (MPA Garching), 
Max-Planck-Institut f\"ur Extraterrestrische Physik (MPE), 
National Astronomical Observatories of China, New Mexico State University, 
New York University, University of Notre Dame, 
Observat\'ario Nacional / MCTI, The Ohio State University, 
Pennsylvania State University, Shanghai Astronomical Observatory, 
United Kingdom Participation Group,
Universidad Nacional Aut\'onoma de M\'exico, University of Arizona, 
University of Colorado Boulder, University of Oxford, University of Portsmouth, 
University of Utah, University of Virginia, University of Washington, University of Wisconsin, 
Vanderbilt University, and Yale University.

\clearpage
\onecolumn
\appendix
\section{Additional Data}

\begin{table}
\centering
\caption{The numerical values used in classifying the H$\alpha$ morphology of each galaxy. When a galaxy possessed more than one of these features, both values were used. For example, a galaxy possessing H$\alpha$ both at the centre and ends of the bar would be given the H$\alpha$ morphology value of 23. }
\label{A1}
\begin{tabular}{c l}
\hline
\hline
\textbf{H$\alpha$ morphology}  & \textbf{Description}   \\
\textbf{value}   &   \\
\hline
   0  &  No H$\alpha$\\
   1  &  H$\alpha$ along the bar\\
   2  &  H$\alpha$ in the centre\\
   3  &  H$\alpha$ at the ends of the bar\\
   4  &  H$\alpha$ in a ring\\
   5  &  H$\alpha$ predominantly outside the bar region in the outer disk\\
   6  &  Unclassifiable - unresolved, or doesn't fit into any of the above categories \\ 
     &  (including H$\alpha$ present, but not associated with the bar regions of the galaxy) \\
   \hline
\end{tabular}
 \end{table}

\begin{table}
\centering
\caption{A list of each H$\alpha$ morphology numerical combination used in this work, and the corresponding H$\alpha$ morphology category it was placed into based on the hierarchy detailed in Section~\ref{classifications}.}
\label{A2}
\begin{tabular}{l l}
\hline
\hline
\textbf{Category}    & \textbf{H$\alpha$ morphology combinations in each category}    \\
\hline
   H$\alpha$ present along the bar  & 1, 12, 123, 1234, 125, 13, 14, 124, 15 \\
   Predominantly central emission &  2, 25 \\
   Prominent ring & 4, 24, 45, 245  \\
   H$\alpha$ at the ends of the bar, or centre and ends  & 3, 23, 35, 235, 234 \\
   No H$\alpha$ detected & 0 \\
   Unclassifiable  & 5, 6 \\
   \hline
\end{tabular}
\end{table}

\footnotesize
\begin{longtable}{l c c c c c c c c c}
\caption{Additional data for all 684 galaxies in the barred galaxy sample. The first ten entries are shown here, and the full table will be available electronically.}\\
\label{A3}\\
\hline
\hline
\textbf{MaNGA}      &   \textbf{Stellar mass}$^{1}$   &   $\rm\textbf{{R}}_{e}^{1}$ ($^{\prime \prime}$)     &   \textbf{H$\alpha$}   &   \textbf{Notes}                &   
\textbf{Arm}    &   \textbf{Stellar bar}    &   \textbf{Stellar bar}   &   \textbf{H$\alpha$ bar}    &   \textbf{H$\alpha$ bar}   \\  
\textbf{plate-ifu}         &  ($\times10^{10}~\rm{M}_{\odot}$)   &        & \textbf{morph.}$^{2}$ & &\textbf{rotation}$^{3}$ &\textbf{length (kpc)} & \textbf{PA} (\textdegree) & \textbf{length (kpc)}  & \textbf{PA} (\textdegree) \\
\hline
 8250-12703    &     0.02           &      15.0   &      1          &      --                   &      3              &      0.3              &      72.2        &      0.9                 &      51.1           \\
  8150-12702    &     0.04           &      12.4   &      5          &      --                   &      --             &      2.1              &      6.3         &      --                  &      --             \\
  8977-12705    &     0.05           &      8.4    &      1          &      --                   &      3              &      2.9              &      116.8       &      3.6                 &      102.2          \\
  8623-9101     &     0.05           &      7.8    &      1          &      --                   &      2              &      1.6              &      33.2        &      0.3                 &      14.4           \\
  8657-12704    &     0.05           &      8.4    &      1          &      --                   &      3              &      3.0              &      97.1        &      3.0                 &      110.5          \\
  8980-12704    &     0.05           &      12.7   &      1          &      --                   &      3              &      1.9              &      160.4       &      --                  &      --             \\
  8461-1901     &     0.06           &      1.9    &      2          &      --                   &      --             &      0.8              &      29.9        &      --                  &      --             \\
  8147-9102     &     0.06           &      7.8    &      3          &      --                   &      --             &      0.8              &      76.8        &      --                  &      --             \\
  8552-3701     &     0.07           &      4.5    &      2          &      --                   &      --             &      2.7              &      112.4       &      --                  &      --             \\
  8713-3701     &     0.07           &      13.5   &      6          &      --                   &      --             &      1.4              &      57.4        &      --                  &      --             \\
  \hline
 \multicolumn{10}{l}{$^{1}$ Elliptical Petrosian photometry values from the NASA Sloan Atlas.}\\
 \multicolumn{10}{l}{$^{2}$ See H$\alpha$ morphology description in Tables~\ref{A1} and ~\ref{A2}}.\\
 \multicolumn{10}{l}{$^{3}$ 1 = clockwise, 2 = anticlockwise, 3 = unsure/ S0, no value = not in the H$\alpha$ along bar subsample.}\\
 \hline
\end{longtable}

\twocolumn

    \bibliographystyle{mnras}
  \bibliography{barcensus}
\end{document}